\documentclass[aip,jap,amsmath,amssymb,preprint]{revtex4-2}

\usepackage{graphicx}
\usepackage{dcolumn}
\usepackage{bm}
\usepackage{amsmath}
\usepackage{url} 
\usepackage{hyperref} 
\usepackage{xcolor}
\usepackage[utf8]{inputenc}
\usepackage[T1]{fontenc}
\usepackage{mathptmx}
\usepackage{etoolbox}

\makeatletter
\def\@email#1#2{%
 \endgroup
 \patchcmd{\titleblock@produce}
  {\frontmatter@RRAPformat}
  {\frontmatter@RRAPformat{\produce@RRAP{*#1\href{mailto:#2}{#2}}}\frontmatter@RRAPformat}
  {}{}
}%
\makeatother
\begin{document}

\title{Effect of intense x-ray free-electron laser transient gratings on the magnetic domain structure of Tm:YIG}

\author{Victor Ukleev}
\email{victor.ukleev@psi.ch, urs.staub@psi.ch}
\thanks{Present address: Helmholtz-Zentrum Berlin f\"ur Materialien und Energie, D-14109 Berlin, Germany}

\author{Max Burian}
\thanks{V. Ukleev and M. Burian contributed equally to this work}
\affiliation{Swiss Light Source, Paul Scherrer Institute, 5232 Villigen PSI, Switzerland}

\author{Sebastian Gliga}
\affiliation{Swiss Light Source, Paul Scherrer Institute, 5232 Villigen PSI, Switzerland}

\author{C.A.F. Vaz}
\affiliation{Swiss Light Source, Paul Scherrer Institute, 5232 Villigen PSI, Switzerland}

\author{Benedikt R\"osner}
\affiliation{Laboratory for Non Linear Optics, Paul Scherrer Institute, 5232 Villigen PSI, Switzerland}

\author{Danny Fainozzi}
\affiliation{Elettra - Sincrotrone Trieste S.C.p.A., 34149 Basovizza, Trieste, Italy}

\author{Gediminas Seniutinas}
\affiliation{Laboratory for X-ray Nanoscience and Technologies, Paul Scherrer Institute, 5232 Villigen PSI, Switzerland}

\author{Adam Kubec}
\affiliation{Laboratory for X-ray Nanoscience and Technologies, Paul Scherrer Institute, 5232 Villigen PSI, Switzerland}

\author{Roman Mankowsky}
\affiliation{SwissFEL, Paul Scherrer Institute, 5232 Villigen PSI, Switzerland}

\author{Henrik T. Lemke}
\affiliation{SwissFEL, Paul Scherrer Institute, 5232 Villigen PSI, Switzerland}

\author{Ethan R. Rosenberg}
\affiliation{Department of Materials Science and Engineering, Massachusetts Institute of Technology, Cambridge, MA 02139, USA}

\author{Caroline A. Ross}
\affiliation{Department of Materials Science and Engineering, Massachusetts Institute of Technology, Cambridge, MA 02139, USA}

\author{Elisabeth M\"uller Gubler}
\affiliation{Electron Microscopy Facility, Paul Scherrer Institute, 5232 Villigen PSI, Switzerland}

\author{Christian David}
\affiliation{Laboratory for X-ray Nanoscience and Technologies, Paul Scherrer Institute, 5232 Villigen PSI, Switzerland}

\author{Cristian Svetina}
\affiliation{SwissFEL, Paul Scherrer Institute, 5232 Villigen PSI, Switzerland}

\author{Urs Staub}
\affiliation{Swiss Light Source, Paul Scherrer Institute, 5232 Villigen PSI, Switzerland}

\date{\today}

\begin{abstract}
Magnetic patterns can be controlled globally using fields or spin polarized currents. In contrast, the local control of the magnetization on the nanometer length scale remains challenging. Here, we demonstrate how magnetic domain patterns in a Tm-doped yttrium iron garnet (Tm:YIG) thin film with perpendicular magnetic anisotropy can be permanently and locally imprinted by high intensity photon pulses of a hard x-ray transient grating (XTG). Micromagnetic simulations provide a qualitative understanding of the observed changes in the orientation of magnetic domains in Tm:YIG and XTG-induced changes. The presented results offer a route for the local manipulation of the magnetic state using hard XTG.
\end{abstract}

\maketitle

\section{Introduction}

The magnetic state of ferromagnetic materials can typically be controlled using magnetic fields or spin-polarized currents. However, these modify the magnetic state globally, and do not allow a targeted local control of magnetization. Such manipulation can be achieved by tailoring the shape and spacing of ferromagnetic nanoelements to achieve the required response \cite{chaurasiya2021comparison,hertel2007ultrafast} to a global field or current. This approach is impractical in extended ferromagnetic samples, where the only means of local control the magnetic state needs to exploit the interaction between the magnetic nanoelements. \cite{wohlhuter2015nanoscale}
Alternatively, optical control of magnetization can be carried out at the sub-micrometer length and femtosecond time scales \cite{lambert2014all,kimel2019writing}. Most of these studies were conducted using visible or infrared light, while rare examples have demonstrated local manipulation of magnetic textures using X-rays \cite{guang2020creating}. Here, we present a route to achieving local control of the magnetic state by using periodic X-ray transient gratings.


X-ray transient gratings (XTG) are formed by interfering coherent beams at the sample to generate spatially periodic excitation patterns. These patterns can equally display temporal structure, e.g., by using periodic laser pulses. XTGs offer unique opportunities to manipulate the structural and electronic properties of materials at the femtosecond timescale down to spatial scales of a few nanometers.\cite{eichler2013laser,bencivenga2019nanoscale,svetina2019towards,rouxel2021hard}
In optical and XUV spectroscopies \cite{eichler2013laser,bencivenga2019nanoscale} splitting and crossing two laser pulses by a set of mirrors is a standard method to generate XTGs; in contrast, this is non-trivial in the soft x-ray and hard x-ray regimes. \cite{cocco2022wavefront} However, recent developments in x-ray free-electron lasers (XFEL) and x-ray optics have enabled an extension of the XTG technique towards hard x-ray energies. \cite{katayama2013femtosecond,svetina2019towards,rouxel2021hard,peters2021hard} 

In the present study we report the successful manipulation of the periodicity and of the spatial orientation of magnetic domains within a thin magnetic film by imprinting a hard x-ray grating with high fluence that causes permanent structural changes of the material, leading to characteristic changes in the magnetic structure.

\begin{figure*}
    \includegraphics[width=1\textwidth]{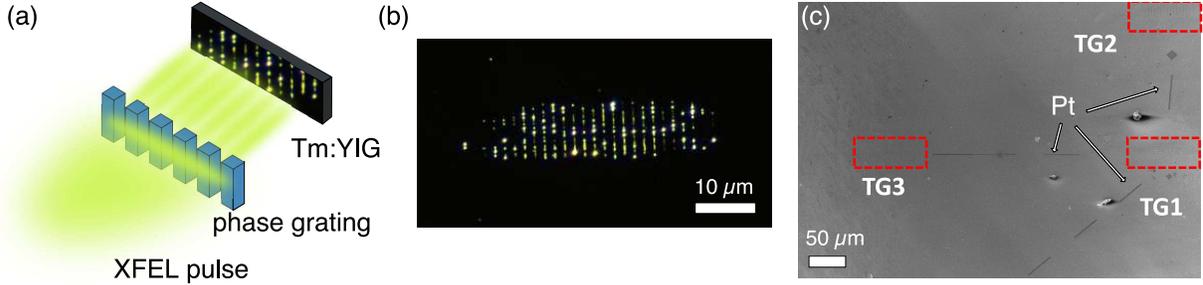}
    \caption{(a) Schematic of the x-ray transient grating experiment. The incoming XFEL beam is diffracted by a transmission phase grating generating an interference pattern (so-called Talbot carpet). The pattern is permanently imprinted on the Tm:YIG sample placed downstream the phase grating. (b) Dark field optical microscopy image of the permanent grating imprinted on the sample by the hard x-ray transient grating with a beam fluence of 70\,mJ/cm$^2$ (marked as TG1 in the panel (c)). (c) Scanning electron microscopy (SEM) image of the sample surface. Pt markers have been deposited to identify different XTG irradiated areas marked as TG1, TG2 and TG3 corresponding to beam fluences of 70\,mJ/cm$^2$, 17.5\,mJ/cm$^2$ and 3,5\,mJ/cm$^2$, respectively.}
    \label{Fig1}
\end{figure*}

\section{Experimental}

A 24\,nm-thick thulium‐substituted yttrium iron garnet  Y$_{0.51}$Tm$_{2.49}$Fe$_5$O$_{12}$ (Tm:YIG) film displaying perpendicular magnetic anisotropy (PMA) was grown on a (111) gadolinium gallium garnet (Gd$_3$Ga$_5$O$_{12}$, GGG) substrate by means of pulsed laser deposition (PLD). Details of the PLD synthesis and sample characterization are given in Refs. \onlinecite{buttner2020thermal,rosenberg2021magnetic}. This composition yields a film with perpendicular magnetic anisotropy lower than that of TmIG but with similar magnetization.

Highly-intense XFEL pulses with 40\,fs duration and 50\,Hz repetition rate were delivered by SwissFEL at the Bernina beamline \cite{ingold2019experimental} to imprint the grating onto the Tm:YIG sample in the same setup as in Ref. \onlinecite{rouxel2021hard}. The energy of the incoming XFEL beam was 7.1\,keV with a bandwidth of ~0.3\%. In total, 1000 XFEL pulses with duration of $\sim40$\,fs each were utilized to permanently imprint the gratings onto the sample. The x-ray beam fluence was controlled by a set of attenuators, varying from 3.5\,mJ/cm$^2$ to 70\,mJ/cm$^2$ at the sample, corresponding to an average of 5\% and 100\% of peak fluence of the full beam, respectively. A schematic illustration of the XTG experiment is shown in Fig. \ref{Fig1}a. The XTG pattern was generated by diffracting the incoming hard x-ray beam on the transmission phase grating. The quasi-one-dimensional grating with a spatial repetition period of $\Lambda = 1650$\,nm was made of polycrystalline chemical-vapour-deposited diamond. Details on the grating fabrication can be found in Refs. \onlinecite{makita2017fabrication,rouxel2021hard}. 

The grating was placed at a distance of 150\,mm upstream of the sample (Fig. \ref{Fig1}a). The sample-to-grating distance and the real-space periodicity of the phase grating $\Lambda$
determine the periodicity of the XTG pattern at the sample position, which can be either smaller or larger than $\Lambda$ depending on the convergence or divergence of the incident photon beam.\cite{rouxel2021hard} In the present case, the period of the phase grating and its distance to the sample were chosen so that the real-space XTG period is of the order of the width of magnetic domains in Tm:YIG, of a few $\mu$m. The dark field image recorded with an optical microscope presented in Fig. \ref{Fig1}b shows the contrast generated by the permanent XTG imprint due to damage of the Tm:YIG film. To simplify the navigation on the sample surface in the experiments, Pt markers have been deposited on Tm:YIG by means of focused ion beam (FIB) as shown in Fig. \ref{Fig1}c.

\begin{figure*}
    \centering
    \includegraphics[width=1\textwidth]{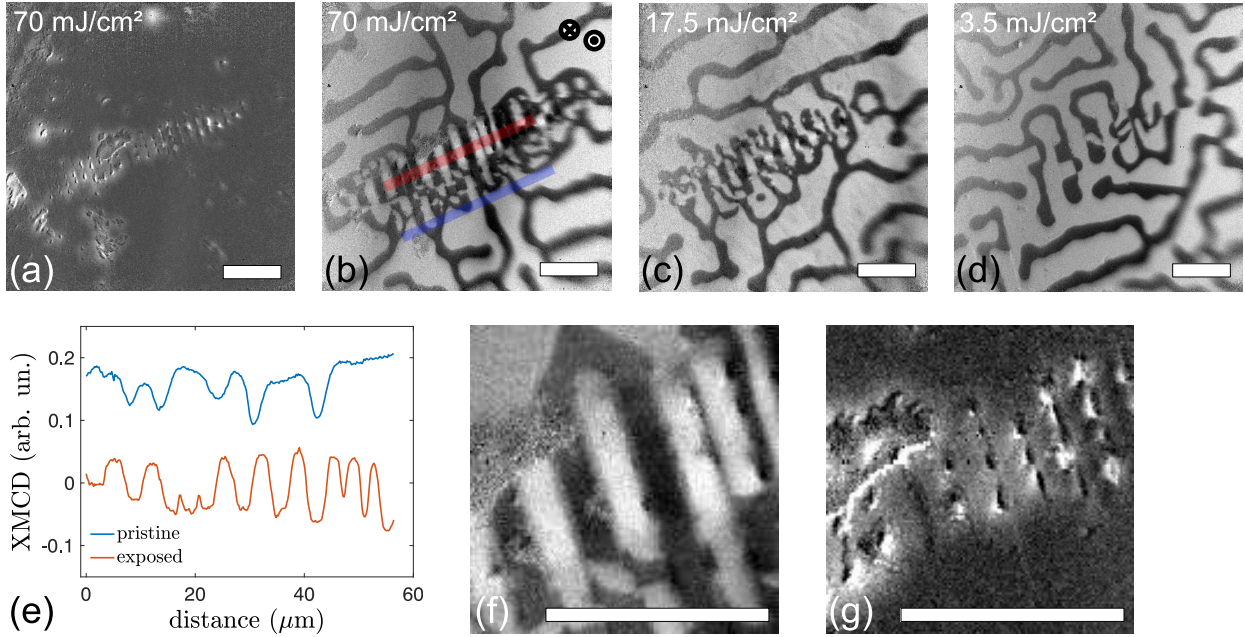}
     \caption{(a) Topography of the sample area exposed with the XTG with the maximum beam fluence of 70\,mJ/cm$^2$ and the corresponding XMCD-PEEM image of the magnetic contrast (b). (c,d) XMCD contrast in the areas of the sample exposed to XTG with fluences of 17.5\,mJ/cm$^2$ and 3.5\,mJ/cm$^2$, respectively. (e) Line profiles of the XMCD intensity taken along the pristine (blue) and XTG-exposed (red) areas of the film, highlighted in the panel (b) by the corresponding colour. The blue curve is shifted by +0.15 for clarity. (f,g) Magnified images of the XMCD and topography contrasts shown in panels (b) and (a), respectively. White scale bars correspond to 20\,$\mu$m.}
    \label{Fig2}
\end{figure*}

\section{Results and discussion}

The magnetic domain structure of the film was measured by photoemission electron microscopy at the SIM beamline \cite{flechsig2010performance} of the Swiss Light Source (PSI, Switzerland) by exploiting the x-ray magnetic circular dichroism effect (XMCD-PEEM). In this technique, x-ray light (here tuned to the Fe $L_3$ absorption edge at 710.6\,eV) uniformly illuminates the sample and the intensity of photoemitted secondary electrons is imaged to obtain local maps of the x-ray absorption of the sample with a spatial resolution down to 50 nm.\cite{vaz2020nanoscale} By averaging PEEM images measured with opposite circular polarizations $C^+$ and $C^-$ one obtains a map of the local electron photoemission, which is very sensitive to the local surface potential and to the sample morphology. The sum $(C^+ + C^-)$ image is shown in Fig. \ref{Fig2}a. The stripes visible in the surface topography correspond well to the optical microscopy result shown in Fig. \ref{Fig1}b and is due to the permanent modification of the material structure induced by the XTG exposure. 

Magnetic XMCD images $(C^+ - C^-) / (C^+ + C^-)$ are shown in Figs. \ref{Fig2}b-d for regions exposed to the XTGs at different fluences. Figure \ref{Fig2}b shows the XMCD contrast for the same area as Fig.\ref{Fig2}a. Domain patterns typical for YIG-based systems with out-of-plane anisotropy \cite{chun2007spin,xia2010investigation,zavislyak2013electric,wang2016evolution,ghising2017stripe,rosenberg2021magnetic} are present in the pristine regions of the sample.  We note that the pristine regions exhibit an asymmetry between the widths of the out-of-plane domains pointing parallel and antiparallel to the film normal (areas of bright and dark contrast). We observe a width $\sim20$\,$\mu$m for the bright domains, as compared to a width of $\sim3$\,$\mu$m for the dark domains (Figs. \ref{Fig2}b-d). We attribute this to the presence of a small magnetic bias field, \cite{gliga2017emergent} or, alternatively, non-zero remanent magnetization of the film.\cite{rosenberg2021magnetic} 
The magnetic domain patterns are clearly different in the irradiated areas. 
For the maximal XTG fluence (Fig. \ref{Fig2}b), the orientation of the magnetic domain stripes is visibly modified and aligned with the regions exposed to the XTG, creating parallel band domains in the irradiated region. In the exposed regions where the XTG fluence is lower, more random domain patterns are observed (Figures \ref{Fig2}c,d). 
In the XTG-imprinted area, the amplitude of the magnetic contrast remains comparable or even stronger to the unexposed one, as seen from the line profiles given in Fig. \ref{Fig2}e. This indicates that the magnitude of the magnetic moment is not reduced by the imprint. Furthermore, the increased contrast is evident from Fig. \ref{Fig2}f which indicates either increase of the magnetization or its tilting towards the sample plane.

In the case of the permanently imprinted gratings with fluences 70\,mJ/cm$^2$ and 17.5\,mJ/cm$^2$ (Figures \ref{Fig2}b,c) the periodicity and the size of magnetic domains differ from those in the pristine area. This is clearly seen in the extracted line profiles of the XMCD-PEEM intensities (Fig. \ref{Fig2}e) in the regions of interest for 70\,mJ/cm$^2$ marked by red and blue lines in Fig. \ref{Fig2}b. 
We find that the average distance between magnetic domains is of $5.3 \pm 1.5$ $\mu$m in the exposed region and of $l = 8.7 \pm 1.5$ $\mu$m in a nearby pristine region, as extracted from the red and blue regions in \ref{Fig2}b, respectively. 


In the XTG irradiated regions, the presence of closely spaced domains is likely due to two processes: 1) local demagnetization, and 2) pinning within the damaged areas. The former explains the varying degrees of modification of the domain pattern as a function of fluence. 
The latter is supported by the fact that domain pinning takes place exactly at the permanently imprinted grating as seen in Figures \ref{Fig2}f,g, which respectively show the magnetic contrast and surface topography in the area exposed to 70\,mJ/cm$^2$ XTG. 
In the case of the intermediate fluence of 17.5\,mJ/cm$^2$, lower pinning contributes to the the more randomly altered magnetic domain pattern, as compared to that at 70\,mJ/cm$^2$. 




To better understand the qualitative changes of the magnetic domain pattern induced by the XTG, the effect of the spatially periodic demagnetization on Tm:YIG system was investigated by micromagnetic simulations.\cite{vansteenkiste2014design} The simulated geometry consisted in a thin film with dimensions of $9000\times9000\times24$ nm$^3$ with a cell size of $6\times6\times6$\,nm$^3$. The exchange stiffness $A_{ex} = 2.3$\,pJ/m, first-order uniaxial out-of-plane anisotropy constant $K_{u1} = 18$\,kJ/m$^3$, and saturation magnetization $M_s = 140$\,kA/m typical for Tm:YIG \cite{buttner2020thermal} were used as material parameters. The Gilbert damping constant was taken to be $\alpha=1$ to accelerate convergence of the simulations. The present simulations did not take into account material damage due to the x-ray irradiation nor ultrafast processes taking place at the sub-ns time scale.

\begin{figure*}
    \centering
    \includegraphics[width=1\textwidth]{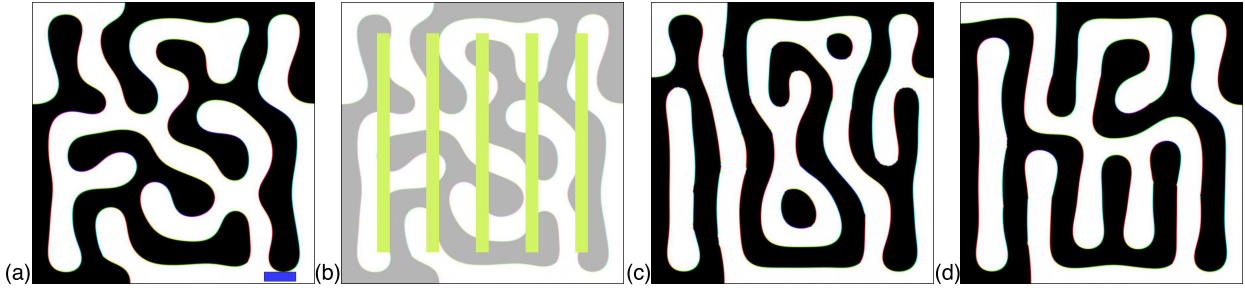}
     \caption{
     Micromagnetic simulations of the XTG effect on the magnetic domain pattern. Starting with a pristine domain pattern (a), the effects of the XTG are simulated by defining regions of the sample (green) where the magnetization is transiently suppressed (b). A modified equilibrium magnetization pattern emerges after transient exposure with (c) 400 nm-wide and (d) 100 nm-wide gratings. The blue scale bar corresponds to 1\,$\mu$m.}
    \label{Fig3}
\end{figure*}

An initial magnetic configuration resulting from the competition between the magnetostatic, exchange and anisotropy energies, and displaying a maze domain was used (Fig. \ref{Fig3}a).
Note that this configuration exhibits equal regions with dark and bright contrasts. Due to the relative low Curie temperature of Tm:YIG ($\sim500$\,K) we anticipate a total quenching of the magnetization in the irradiated areas due to the local heating above $T_\textrm{C}$ by the laser pulses. \cite{wang2012femtosecond} The effects of the XTG were simulated by assuming the total quenching of the magnetization due to the transient grating, i.e., by locally setting $A_{ex}$, $M_s$ and $K_u$ to zero within rectangular regions (Fig. \ref{Fig3}b). The width of the modified regions was 400\,nm. This quenching of the magnetization leads to a redistribution of the magnetization along the boundaries of the exposed regions. The magnetization is subsequently restored with the same material parameters, giving rise to a modified pattern (Fig. \ref{Fig3}c) made of band domains that closely match the position of the transient gratings, reflecting the experiments. 

To explain the observed change in domain periodicity in the irradiated regions, we suggest that magnetization pinning occurs in regions where the material's magnetic parameters have been strongly altered. Although the surrounding magnetization may have recovered, the persistent pinning may be due to physical changes of the sample. 

We expect that laser-induced heating is the main source of the magnetization quenching in the XTG-exposed regions. Higher hard x-ray fluences heat the irradiated areas stronger, with the heat being dissipated to a larger area around the XTG-exposed stripes via phonons. \cite{rouxel2021hard} Reduction of the irradiation dose can be qualitatively mimicked by narrowing the width of demagnetized regions to 100\,nm in the simulation. The result is a more disordered imprinted pattern (Fig. \ref{Fig3}d), which does not lead to the formation of parallel band domains, in qualitative agreement with the patterns observed in the sample areas exposed with XTGs with the beam fluences of 17.5\,mJ/cm$^2$ and 3.5\,mJ/cm$^2$ (Figures \ref{Fig2}c,d). Moreover, the simulations show that in the absence of physical pinning, the imprint does not affect the width of magnetic domains which is determined by the competition between the exchange, magnetostatic and anisotropy energies. Instead, the grating geometry affects the orientation of the domains in the irradiated area. Despite the simplicity of the model, the simulations reproduce the main effects of the XTG. The fact that the simulations display features that are not experimentally observed (bubble states in Fig. \ref{Fig3}c) or do not reproduce certain experimentally-observed features (such as domains that are perfectly parallel to the XTG) points to the role played by physical defects and changes in material parameters beside the demagnetization. Therefore, we discuss a few possible mechanisms that may affect the magnetic structure in the XTG-exposed regions.

Indeed, aside from structural and morphological changes in the sample, the domain wall pinning could be induced by the local modification of uniaxial anisotropy in the irradiated regions. 
It is also known that while maze patterns form in the presence of uniaxial out-of-plane anisotropy, the formation of parallel band domains requires the presence of additional, secondary anisotropy contributions superimposed to the fundamental out-of-plane anisotropy.\cite{hubert1974effect} Moreover, the stronger contrast in the exposed areas seen in Fig. \ref{Fig2}b hints towards the in-plane magnetization component.

While intense optical pulses and ion irradiation can result in structural and associated magnetic changes in iron garnets, \cite{flanders1971photoinduced,lems1970light,holtwijk1970light,haisma1972direct,gyorgy1971irreversible,hansen1972anisotropy,metselaar1973stoichiometry} the effect of hard x-rays is less explored, although x-ray induced damage has so far also been observed in some other magnetic oxides \cite{kiryukhin1997x,kiryukhin1999x,garganourakis2012imprinting,yamasaki2015x}. 
Therefore, the exact mechanism of the magnetic domain change, as well as other possible scenarios, would require deeper follow up studies.

\section{Conclusion}

In conclusion, we have investigated  magnetic domain structures imprinted in a Tm:YIG PMA film by an XFEL hard x-ray transient grating. By measuring the resulting magnetic patterns with XMCD-PEEM, we observed modifications of the domain patterns in the exposed regions that correlate with permanent changes in the sample. Particularly, the observed decrease of the magnetic domain spacing and their orientation suggests a pinning of the domain walls to the defects imprinted by the x-rays. XFELs allow the generation of gratings with periods down to a few nanometres and durations of tens of femtoseconds, offering a possible route for the ultrafast manipulation of magnetic structures by x-rays down to the nanometer scale in suitable materials. Although the currently studied material only supports micrometer-scale magnetic domains, further investigations could lead to promising pathways to imprint magnetic textures, such as bubble domains or topological magnetic skyrmions of a smaller size.\cite{gerlinger2021application,buttner2021observation} Taking advantage of the periodic spatial patterns induced by radiation as used to produce the XTG, these textures could also be arranged into artificial arrays using x-ray gratings with a custom shape, imprinting for example, one-dimensional chains, or two-dimensional hexagonal or square lattices. The XTG approach is more technologically promising than the generation of magnetic skyrmions by using a focused x-ray beam.\cite{guang2020creating} Furthermore, our study extends the XTG approach towards the hard x-ray range, allowing one to manipulate and probe bulk specimens and reach resonant edges of a broad range of elements.\\ 

\section*{Acknowledgments}

This study was supported by the Swiss National Science Foundation (SNSF), grant no. 200021\_165550/1 and no. 200021\_169017 as well as the SNSF National Centers of Competence in Research in Molecular Ultrafast Science and Technology (NCCR MUST-No. 51NF40-183615). C.A.R. and E.R. acknowledge support of SMART, an nCORE Center of the Semiconductor Research Corporation. Part of this work was performed at the Surface/Interface Microscopy (SIM) beamline of the Swiss Light Source (SLS), and at the Bernina beamline of SwissFEL, both at the Paul Scherrer Institut (PSI), Villigen, Switzerland.

\section*{Author Declarations}
\subsection*{Conflict of Interest}
The authors have no conflicts to disclose.
\subsection*{Author Contributions}
V.U. and M.B. analyzed the data and wrote the paper, M.B., C.A.F.V., B.R., D.F., G.S., R.M., H.T.L., C.S., U.S. performed XFEL and PEEM experiments, E.R and C.A.R. synthesized the sample, C.A.F.V., M.B. and E.A.M. prepared the sample for PEEM and characterized it with scanning electron microscopy, C.D. prepared gratings, S.G. performed micromagnetic simulations, M.B., C.D., C.S. and U.S. jointly conceived the project.

\section*{Data Availability Statement}

The data is available from Zenodo repository (Ref. \onlinecite{data2022psi}).

\nocite{*}
\bibliography{aipsamp}

\end{document}